\documentclass[12pt]{iopart}
\usepackage{iopams}
\usepackage{graphicx}
\usepackage[mathlines]{lineno}% Enable numbering of text and display math
\begin{document}

\title[Evanescent modes and level repulsion in SCW]{Theoretical and experimental evidence of level repulsion states and evanescent modes in sonic crystal stubbed waveguides}

\author{V. Romero-Garc\'ia$^1$, J.O. Vasseur$^2$, L.M. Garcia-Raffi$^3$  and A.C. Hladky-Hennion$^2$.}

\address{$^1$ Instituto de Investigaci\'on para la Gesti\'on Integrada de zonas Costeras, Universidad Polit\'ecnica de Valencia, Paranimf 1, 46730, Gandia, Spain}
\address{$^2$ Institut d'Electronique, de Micro\'electronique et de Nanotechnologie,  UMR CNRS 8520, Cit\'e Scientifique, F-59652 Villeneuve d'Ascq C\'edex, France}
\address{$^3$  Instituto Universitario de Matem\'atica Pura y Aplicada, Universidad Polit\'ecnica de Valencia, Camino de Vera s/n, 46022 Valencia}
\ead{virogar1@gmail.com, http://www.personales.upv.es/virogar1}

\begin{abstract}
The complex band structures calculated using the Extended Plane Wave Expansion (EPWE) reveal the presence of evanescent modes in periodic systems, never predicted by the classical $\omega(\vec{k})$ methods, providing novel interpretations of several phenomena as well as a complete picture of the system. In this work we theoretically and experimentally observe that in the ranges of frequencies where a deaf band is traditionally predicted, an evanescent mode with the excitable symmetry appears changing drastically the interpretation of the transmission properties. On the other hand, the simplicity of the sonic crystals in which only the longitudinal polarization can be excited, is used to interpret, without loss of generality, the level repulsion between symmetric and antisymmetric bands in sonic crystals as the presence of an evanescent mode connecting both repelled bands. These evanescent modes, obtained using EPWE, explain both the attenuation produced in this range of frequencies and the transfer of symmetry from one band to the other in good agreement with both experimental results and multiple scattering predictions. Thus, the evanescent properties of the periodic system have been revealed necessary for the design of new acoustic and electromagnetic applications based on periodicity.
\end{abstract}

%\keywords{Phononic crystal, Sonic Crystal, Complex band structures, Evanescent modes, Waveguides}

\maketitle

\section{Introduction}
Since the pioneering works of Yablonovitch \cite{Yablonovitch87, Yablonovitch89} and John \cite{John87} simultaneously discovered the possibilities to control the flow of light in periodic distribution of dielectric materials, a drastic increasing interest appeared in the analogous structures to control both the elastic and acoustic waves using the well-known Phononic \cite{Sigalas93, Kushwaha93} or Sonic Crystals \cite{Martinez95, Kushwaha97, Sanchez98, Robertson98} respectively. Phononic Crystals are periodic distributions of two elastic materials, however in the case that one of these materials is a fluid the system is called Sonic Crystal (SC). In the past twenty years the exploitation of the particular dispersion relation of these structures has revealed very interesting physical properties \cite{Sigalas93,Kushwaha93,Martinez95,Kushwaha94,Kushwaha97,Sanchez98,Robertson98,Sigalas97,Sigalas98,Caballero99,Tanaka00,Wu01,Cervera02,Wu03,Khelif03,Khelif04,Yang04,Feng05,Li05,Sigalas05,Torrent06a,Tanaka07,Perez07,Espinosa07,Vasseur08,Vasseur10,Romero10a,Romero10b,Romero10c,Romero11}. The existence of ranges of frequencies, called Bandgaps (BGs), in which non propagating modes can be excited in the system, has been observed in a wide range of frequencies due to the scalability of the systems \cite{Caballero99, Tanaka00, Vasseur08}. The presence of the BG has motivated the analysis of both the localized states in point defects \cite{Sigalas97,Sigalas98, Wu01, Wu03, Li05} and the guidance of waves thorough a linear defect constituting a waveguide in the periodic system \cite{Khelif03, Khelif04, Tanaka07}. However, other properties of the dispersion relation have been used to control the wave propagation through these periodic structures. On one hand, the linearity of the dispersion curves at subwavelength regimes has motivated the application of homogenization theories in order to define effective parameters of the system \cite{Cervera02, Torrent06a}. On the other hand, the curvature of the isofrequency contours has been used to control the spatial dispersion of the waves inside the periodic structures in order to obtain both self-collimating effect \cite{Perez07, Espinosa07} and negative refraction \cite{Yang04, Feng05}.

Recently, the possibility to control the evanescent properties in periodic composites has shown several interesting possibilities in both photonic \cite{Li07, Dong11} and phononic \cite{Sukhovich09, Robillard11, Zhu11} crystals. Imaging with super resolution,\cite{Pendry00, Fang05} that is, overcoming the diffraction limit can be obtained by restoring all the evanescent components of a near-field image. This can be achieved by the coupling of the evanescent modes with other mechanism leading to their amplification in order to successfully transport the information carried by the evanescent waves through the system. In this way, the evanescent properties of the periodic system have been revealed necessary for the design of new acoustic and electromagnetic applications.

The dispersion relation of these periodic structures has been traditionally obtained using $\omega(\vec{k})$ methods, for instance, the plane wave expansion (PWE) \cite{Kushwaha94}. %Making use of the periodicity of the structure, both Fourier series and Bloch's theorem can be applied to define an eigenvalue problem. The solution of this eigenvalue problem is the dispersion relation $\omega_n(\vec{k})$, where $\omega$ is the angular frequency of the wave (eigenvalue), $\vec{k}$ is the wave vector inside the periodic structure or the Bloch's vector and $n$ is the band number. It is worth noting that the dispersion relation (or band structure) is obtained iteratively for each $\vec{k}$ belonging to the borders of the first irreducible Brillouin zone. For each $\vec{k}$, $n$ eigenvalues $\omega_n$ can be obtained, in such a way $n$ propagating bands can be represented. 
 %In the 
 Using the PWE method, the BGs can be defined as ranges of frequencies in which no modes are excited inside the periodic structure. However, BGs should be defined as ranges of frequencies where only evanescent modes can be excited \cite{Brillouin, Joannopoulos08}. Recent works have shown in both regimes, acoustic \cite{Romero10b} and elastic \cite{Laude09}, that an extension of the PWE method, called the Extended Plane Wave Expansion (EPWE), is an efficient tool for analyzing the evanescent properties of the periodic systems. This methodology inverts the $\omega(\vec{k})$ problem to the $k(\omega)$ problem by fixing both the angular frequency $\omega$ and the direction of incidence $\vec{\alpha}$ ($\vec{k}=k\vec{\alpha}$), $k$ being the eigenvalue of the problem. %Experimental measurements \cite{Romero10b}  of the evanescence of the modes in the BG are in very good agreement with the predictions of the complex band structures obtained using the EPWE ($k_n(\omega)$). 
  The supercell approximation has been introduced in the EPWE to theoretically characterize the evanescence of both single \cite{Romero10a} and double \cite{Romero10c} point defects. The evanescent decay predicted by EPWE for both complete and defective SC is in very good agreement with the experimental measurements \cite{Romero10a, Romero10b, Romero10c}. % The exploitation of the evanescent properties of periodic systems has been recently used to design flat lenses based on phononic crystals showing images with super resolution. \cite{Sukhovich09,Zhu11}
   The picture of the propagation properties of the periodic systems provided by the EPWE method improves and completes that given by the PWE. %Up to the best of our knowledge, both the band structure and the transmission problems should be simultaneously considered to characterize the propagation properties of the systems. However, the information provided by the complex band structure ($k_n(\omega)$ methods) about both the propagating and the non-propagating (evanescent) modes could be used to characterize the complete properties of the system.

The goals of the paper are two-folds. First we observe the evidence of evanescent modes in a sonic crystal waveguide (SCW) using the EPWE with supercell approximation. Second, we analyze the evanescent properties of stubbed SCW. The presence of the stub, breaks the symmetry of the system and fundamental phenomena can be studied. The antisymmetry of the system introduces several splittings in the band structure characterized by an interchange of the properties between the repelled bands. In this work we study the nature of the level repulsion in two-dimensional periodic systems making use of evanescent waves.

We find the evanescent modes of fundamental interest for the correct understanding of the wave propagation properties in both linear waveguides and in general structures based on periodicity. We use a SCW because it allows us to obtain high precision measurements inside the periodic structure where the evanescent properties are excited. As it has been previously shown \cite{Khelif02}, the interest of SCW stands on the effects due to the periodicity that can not exist in rigid walls waveguides. Due to the linear defect in the SCW interesting physics appear in the BG. In this work we theoretically and experimentally observe that for the \textit{deaf} bands \cite{Sanchez98, Rubio99, Hsiao07}, where the real part of the complex band structure represents antisymmetric modes, evanescent modes with the excitable symmetry can appear in the waveguide. In contrast to the classical interpretation of the attenuation in a deaf band, based on geometric arguments, in this work we find that the attenuation is governed by a decay rate related with the imaginary part of the complex band structures obtained using EPWE. Independently, we have obtained the same value of the decay rate using other methodologies, as the Multiple Scattering Theory (MST) \cite{martin06, linton01, Chen01, Kafesaki99}, in very good agreement with the experiments. %On the other hand, evanescent modes have been also observed in stubbed waveguides interacting with the stub resonances. 
 
On the other hand, the mode conversion in phononic crystals thin slab has been recently observed \cite{Chen08}. This conversion depends on the symmetry of the system with respect to the direction of the incident wave. In these systems, the excited modes either couple or not with other ones, being the coupling observed by both a splitting between bands in the band structure and a transfer of the symmetry from one band to the other. In the case in which no coupling occurs, the bands simply crosses between them. Bavencoffe \cite{Bavencoffe09b} established a link between the attenuation of the ultrasonic wave observed in the case of a limited grating and the values of the imaginary part of the wave number in a stop band computed for an infinite grating. In other works, Achaoui \textit{et al.} \cite{Achaoui10}, have recently observed in phononic crystals with some freedom of anisotropy that when a band mostly polarized in-plane is close to a band mostly polarized out-of-plane, a phenomenon of repelling can occur and in some instances it introduces a local band gap. Moreover, this interaction is accompanied by a transfer of the polarization state from one band to the other. These levels repulsion which avoids crossing in the distribution of eigenvalues is well known within the physics community \cite{Dembowski01, Philipp00, Lee03}. However, there are only few works in the literature that analyzes and discusses this phenomenon in two-dimensional phononic crystals \cite{Wu04,Chen08,Achaoui10}.

The simplicity of the SC in which only the longitudinal polarization can be excited, is used to observe without loss of generality the level repulsion between symmetric and antisymmetric bands. In this work we show that the level repulsion between the symmetric and antisymmetric modes is the result of the presence of an evanescent mode. These evanescent modes explain both the attenuation produced in this range of frequencies and the transfer of symmetry from one band to the other. 

The work is organized as follows. First of all both the theoretical techniques and the experimental set up are shown in Section \ref{sec:setup}. In Section \ref{sec:waveguide} we present the classification of the several evanescent modes predicted by the EPWE showing the main characteristics of the complex band structures of SCW. The evanescent behaviour of modes in SCW is also analyzed both analytically using MST and experimentally. Section \ref{sec:coupling} analyses the stubbed SCW. The evanescent bands, non predicted by the traditional PWE, are very relevant in these systems. The interaction of the evanescent modes with cavity modes in the waveguide is shown in Section \ref{sec:waveguide_coupling}. In Section \ref{sec:level_repulsion} the stubbed SCW are used to show the effect of the level repulsion in periodic systems, showing both the local bandgaps and the symmetry transfer between symmetric and antisymmetric bands due to the presence of evanescent modes. The concluding remarks of the work will be shown in Section \ref{sec:conclusions}.

\section{Theoretical techniques and experimental set up}
\label{sec:setup}
The propagation properties of periodic materials can be analyzed solving for both the eigenvalue and the scattering problems. The former can be used to obtain the dispersion relation of a system whereas the latter can be used to obtain the scattering of waves in finite structures. Both methodologies are independently used in this work to analyze the properties of an acoustic waveguide. The eigenvalue problem $k(\omega)$ is solved in this work using the EPWE with supercell approximation \cite{Romero10c}. To compare the results with the classical $\omega(\vec{k})$ some calculations using the PWE method \cite{Kushwaha94} will be shown. To solve the scattering problem we use the MST \cite{martin06, linton01, Chen01, Kafesaki99}, which is a self-consistent method used to analyze the scattered field of an arrangement of scatterers considering all the orders of scattering.

\begin{figure}[hbt]
\begin{center}
\includegraphics[width=85mm]{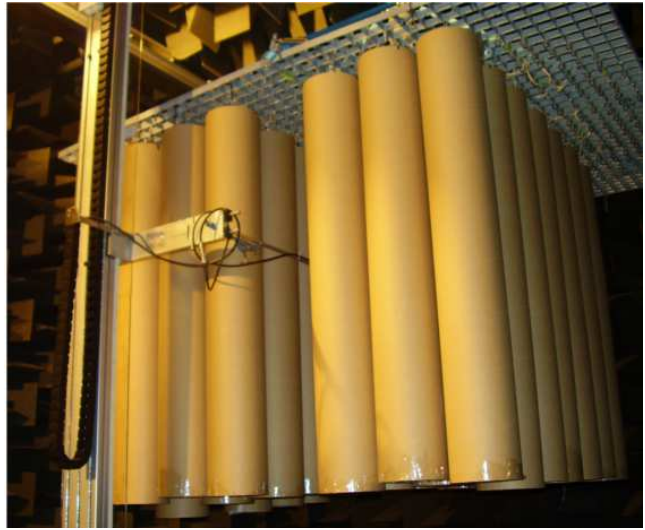}\\
\end{center} %\begin{quote}
\caption{(Color online) Experimental set up. Picture of the frame in which the SCW has been hung and the  complete structure with $7\times7$ rows in which the central one is removed in order to produce the waveguide. %(b) Detailed picture of the microphone inside the SCW. 
}
 \label{fig:set_up}
%\end{quote} 
\end{figure}

The experiments showed in this work have been performed in an echo-free chamber sized $8\times 6\times 3$ m$^3$. To obtain the
experimental dependence of the pressure all along the SCW, we
measured the pressure field inside the guide. We build a finite SC hunging the rigid scatterers on a periodic frame (Figure \ref{fig:set_up}) and  the SCW is generated removing the central row of the complete structure. We notice that this system avoids the excess attenuation effect \cite{Romero11}.
The finite 2D SC is made of 7$\times$7 rigid cylinders arranged as a square array of lattice constant $a$. The radius of the cylinders used in the experiments is $r=9$ cm, and the lattice
constant of the SC is $a=19$ cm. With these parameters, the filling fraction of the finite SC
is $ff=\frac{\pi r^2}{a^2}\simeq0.71$. The dimensions of the system are large enough for the microphone to be placed inside the waveguide. The
microphone used is a prepolarized free-field 1/2" Type $4189$
B\&K. The diameter of the microphone is $1.32$ cm, which is
approximately $0.07a$, therefore a low level of influence over the pressure field measured is expected. The source is placed 1 m away from the SCW launching white noise.

In this work a 3D computer-controlled automatic positioning system together with an automatized acquisition system, called 3DReAMS (3D Robotized
e-Acoustic Measurement System), has been used to obtain the pressure field inside the waveguide. The 3DReAMS system is capable of sweeping the microphone through a 3D grid of measuring points located at any trajectory inside the echo-free chamber.
The robot was controlled by the motion controller card of National Instruments PCI 7334. We analyzed the acoustic field inside the guide of both complete SCW and stubbed SCW by moving the microphone in steps of $1$ cm.

\section{Propagating and evanescent modes in Sonic Crystals Waveguides}
\label{sec:waveguide}
Consider a periodic arrangement of rigid scatterers embedded in air ($\rho_{air}=1.29$ kg/m$^3$, c=340 m/s) with square periodicity which lattice constant is $a$ and the filling fraction $ff=\pi r^2/a^2=0.717$. All the physical parameters presented in the work will be shown in reduced magnitudes: reduced frequencies ($\Psi=\nu a/c_{air}$, $\nu$ is the frequency) and reduced Bloch vector ($K=ka/(2\pi)$). The band structure for this periodic array is shown in Figure \ref{fig:IL_MST}a. The bandgap of the complete SC (without linear defect) is ranged between $\Psi_1=0.33$ and $\Psi_2=0.8$. In this Section we analyze the evanescent and propagating properties of a SCW generated by removing the central row of the complete structure. EPWE shows novel properties of the guided and non-guided modes in the waveguide that would not be observed otherwise using PWE.

\subsection{Preliminary analysis: Transmission properties of a Sonic Crystal Waveguide. Motivation}

\begin{figure}[hbt]
\begin{center}
\includegraphics[width=120mm,height=70mm,angle=0]{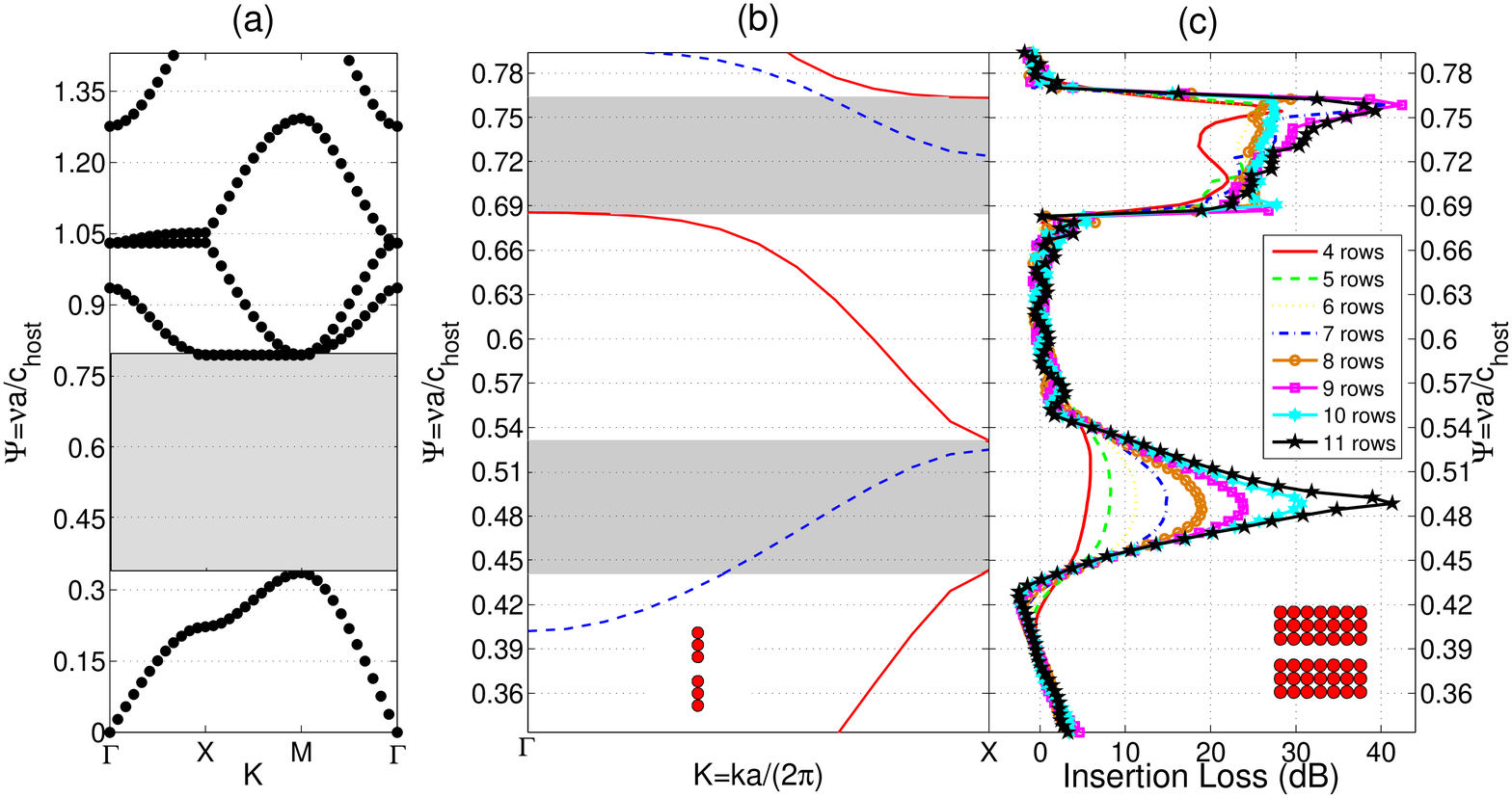}
\end{center} %\begin{quote}
\caption{(Color online) Study of the transmission properties of the SC and SCW. (a) Bands structure for the complete SC made of rigid scatterers embedded in air with square periodicity which lattice constant is $a$ and the filling fraction is $ff=\pi r^2/a^2=0.717$. Grey area represents the bandgap. (b) Bands structure of the SCW at $\Gamma$X direction calculated using PWE (1105 plane waves) for the frequencies in the bandgap of the complete structure. Grey areas represent non guiding bands. Red continuous (Blue dashed) lines represent symmetric (antisymmetric) bands. (c) IL prediction calculated using MST. The different lines represent the IL predictions for SCW with several number of rows (see legend).}
 \label{fig:IL_MST}
%\end{quote} 
\end{figure}

The generated SCW is completely equivalent to that used in a previous work \cite{Vasseur05}. Figure \ref{fig:IL_MST}b shows the band structure along the $\Gamma$X direction for the SCW for these frequencies. The band structure has been calculated using the PWE with the supercell approximation. 1105 plane waves have been used for the calculations. The supercell is shown in the inset of Figure \ref{fig:IL_MST}b.%, contains 7 unit cells and the scatterer of the central one has been removed in order to obtain the SCW once the supercell is periodically placed in the space, giving enough separation between the waveguides to avoid coupling between them.

Because of the linear defect, the band structure in Figure \ref{fig:IL_MST}b shows five guiding bands in the bandgap. The red continuous and blue dashed lines in Figure \ref{fig:IL_MST}b correspond to localized modes in the straight waveguide. The Fourier transform of the eigenvectors yields the pressure field inside the SCW. Analyzing this field \cite{Vasseur05}, one can observe that the modes corresponding to the red continuous lines produce symmetric acoustic fields with respect to main direction of symmetry $\Gamma$X, whereas the acoustic fields corresponding to the blue dashed lines are antisymmetric. Thus the symmetric modes can be excited by a longitudinal incident wave and the antisymmetric modes will not contribute to the transmission because they cannot be excited. The later modes are usually called \textit{deaf} bands and they have been analyzed in the literature \cite{Sanchez98, Rubio99, Hsiao07}.

From the previous discussion it is expected that between symmetric guiding bands should appear a dip of transmission (grey areas in Figure \ref{fig:IL_MST}b). In order to observe this property, one should additionally solve for the scattering problem. We analyze several SCW with different number of rows using MST. We evaluate the Insertion Loss (IL) \cite{Sanchez98}, which corresponds to the attenuation spectrum evaluating the difference between the sound level recorded in the same point without and with the SCW. The comparison between the band structure and the IL spectrum shown in Figure \ref{fig:IL_MST}c indicates that the symmetric bands (red continuous lines in Figure \ref{fig:IL_MST}b) contribute to the transmission  whereas, as it was expected, between this symmetric bands, attenuation peaks appear in the IL spectrum because of the antisymmetric modes (blue dashed lines) cannot be excited. 

Figure \ref{fig:IL_MST}c clearly shows the two attenuation peaks produced by the non excited antisymmetric bands in the waveguide for a high number of rows. However, in the case of low number of rows the attenuation level due to the first antisymmetric band is very low in comparison with the attenuation level of the second peak (red continuous line in Figure \ref{fig:IL_MST}c). Moreover, the attenuation level of the first attenuation peak increases with the number of rows whereas the second peak seems more stable. Thus several fundamental questions reach from this plot. First of all, if band structures predict stop bands for these frequencies (grey areas), what kind of modes can propagate depending on the length of the waveguide? Why does the second antisymmetric mode present larger and more stable attenuation level than the first one? The answers of these questions are related to evanescent waves and cannot be obtained from the results of either the PWE nor the MST. A more accurate approach should be used to understand the physics of these systems.

\subsection{Complex band structures in Sonic Crystals Waveguides}

Brillouin \cite{Brillouin} revealed that waves in the bandgap present evanescent behaviour, i.e., they should be characterized by a complex value of the wavevector in such a way the amplitude of the mode should be modulated by an exponential with negative exponent. Therefore the mode should decay as it penetrates in the crystal.  The elegant and intuitive explanation of Joannopoulos \cite{Joannopoulos08} revealed that the decay rate should be greater for frequencies closer to the center of the BG. Classical methods do not show this kind of properties because they interpret the BG as ranges of frequencies where no modes can be excited.

\begin{figure*}[hbt]
\begin{center}
\includegraphics[width=150mm,height=70mm,angle=0]{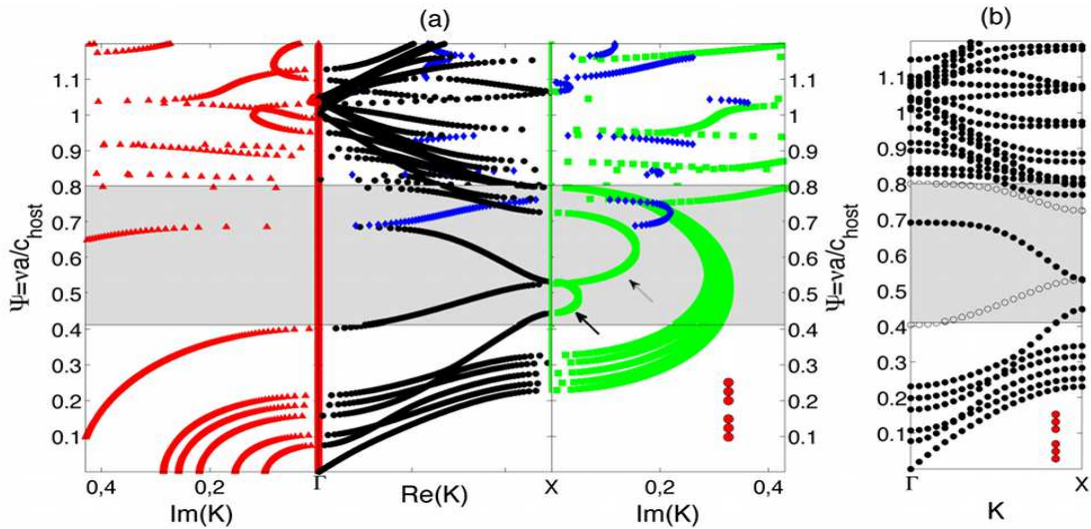}
\end{center} %\begin{quote}
\caption{(Color online) Analysis of the band structure of the SCW. Grey area represents the bandgap. Reduced magnitudes have been used in the plots. (a) Calculation using EPWE. Red triangles represent modes with $Im(K)\neq0$ and $Re(K)=0$, black circles represent modes with $Re(K)\neq0$ and $Im(K)=0$, green squares show modes with $Re(K)=0.5$ and $Im(K)\neq0$ and blue diamonds represent modes with $0<Re(K)<0.5$ and $Im(K)\neq0$. (b) Bands structure calculated using PWE. Inside the bandgap, open circles (black circles) represent antisymmetric (symmetric) modes. The supercell used in both cases are shown in the insets of (a) and (b).}
 \label{fig:Comparison_PWE_EPWE}
%\end{quote} 
\end{figure*}

The Joannopoulos predictions have been recently observed both theoretically \cite{Laude09, Romero10b} and experimentally \cite{Romero10b, Romero10a} for SCs. %The complex band structures obtained using the $k(\omega)$ methods, for instance EPWE, show that the bandgaps are ranges of frequencies where only evanescent modes can be excited inside the SC, i.e., they are characterized by a complex value of $k$. %On the other hand, one of the most striking characteristic of the complex band structure is that each complex band presents a continuous frequency dependence changing from propagating to evanescent behaviour depending on the frequency. Then, consecutive bands are connected with a complex loop in the imaginary part that shows the evanescent behaviour of the modes in the pseudogaps at each direction of symmetry. This mechanism is repeated for higher frequencies and for each complex band in such a way the overall number of bands at any frequency is globally preserved  \cite{Laude09, Romero10b}. 
Motivated by the properties of the complex band structures and by the interesting questions related to evanescent modes presented in the previous Section, we study the evanescent and guiding properties of SCW. %This system shows the relevance to consider the excitation of evanescent modes. 
The conclusions of this work are general and can be applied to other kind of periodic systems.

Due to the freedom of the eigenvalues in the EPWE method (complex valued $k$), a more general band structure than the classical ones are obtained \cite{Romero10b, Laude09}. In this Section we classify the propagating and evanescent modes appearing in the complex band structure resulting from the EPWE method. The classification shown in this work, inspired by the works of Bavencoffe \textit{et al.} \cite{Bavencoffe09b, Bavencoffe09a}, helps to a better understanding of the complex band structures. The modes are classified following the next restrictions: ($i$) The classical band structures correspond to modes characterized by values of $Re(K)$ in the Brillouin zone and $Im(K)=0$. Modes with these properties are shown in this work with black filled circles. ($ii$) The modes characterized by $Im(K)\geq0$ and $Re(K)=0$ are shown with red filled triangles. These modes represent connections between propagating bands in the $\Gamma$ point. ($iii$) The modes characterized by $Im(K)\geq0$ and $Re(K)=1/2$ (respectively, $Re(K)=1/\sqrt{2}$) are shown with green filled squares. These modes represent connections between propagating bands in the $X$ (respectively, M) point. ($iv$) The modes with $Re(K)$ in the first Brillouin zone but with $Im(K)\neq0$ are shown in blue filled diamonds. These modes belong to evanescent connecting bands between bands with the same symmetry of the acoustic field. For this last group we constrain the plots for modes with $Im(K)\leq0.4$.

Following the previous classification we have plotted the complex band structure calculated using the EPWE with supercell approximation for the SCW we are dealing with. The convergence of all the calculations have been carefully analyzed selecting the adequate number of plane waves for each case. 1369 plane waves have been used for the calculation. Figure \ref{fig:Comparison_PWE_EPWE}a shows the complex band structure for the SCW at $\Gamma$X direction. The values of the real part of the wave vector (central panel of Figure \ref{fig:Comparison_PWE_EPWE}a) are restricted to the first Brillouin zone, whereas the imaginary part (right and left panels of Figures \ref{fig:Comparison_PWE_EPWE}a) is arbitrarily displayed. It is worth noting that due to the frequency discretization to solve the EPWE, it is difficult to find flat bands in the band structure. Thus, the frequency step used in this work has been specifically selected for each calculation. %Figure \ref{fig:Comparison_PWE_EPWE}a shows the band structure calculated using PWE with supercell approximation (1105 plane waves). In the inset of both Figures one can see the supercell used for the calculation. 
The grey area shows the range of frequencies of the bandgap of the perfect complete SC. The supercell used for the calculations is shown in the inset of Figure \ref{fig:Comparison_PWE_EPWE}a.

\begin{figure*}[hbt]
\begin{center}
\includegraphics[width=165mm,height=90mm,angle=0]{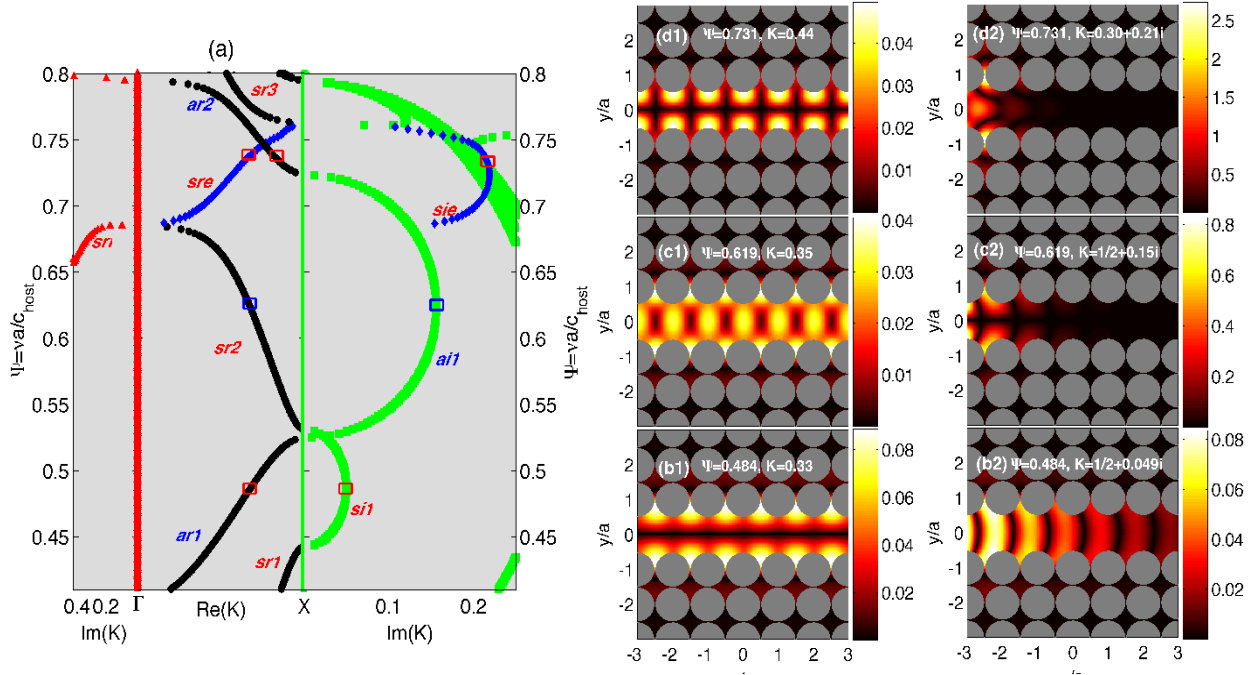}
\end{center} %\begin{quote}
\caption{(Color online) Analysis of the propagating and evanescent modes of the SCW predicted by EPWE (modes at $\Psi=0.484$ $\Psi_2=0.619$ and $\Psi_3=0.731$). (a) Complex bands in the band gap region. (b1)-(d1) Modulus of the acoustic field for the real eigenvalues. (b2)-(d2) Modulus of the acoustic field for the complex eigenvalues.}
 \label{fig:FFTfields_EPWE}
%\end{quote} 
\end{figure*}

Comparing the band structures calculated using EPWE and PWE in Figures \ref{fig:Comparison_PWE_EPWE}a and \ref{fig:Comparison_PWE_EPWE}b respectively, one can observe that EPWE reproduces the same guided modes in the bandgap as in the case of PWE. However two important characteristics should be noted. First, the complex band structure shows that two consecutive symmetric (respectively, antisymmetric) real bands are connected by a symmetric (respectively, antisymmetric) imaginary band indicated with black continuous (respectively, dotted) arrow in Figure \ref{fig:Comparison_PWE_EPWE}a. Therefore the complex connections preserve the symmetry of the bands which are connecting. Second, as a consequence of the connections and due to the folding of the bands, an additional band not predicted by PWE appears in the bandgap (see real and imaginary bands represented with blue diamonds). Let us analyze in detail these two features in the next lines.

At $\Gamma$ or X points where the folding of the bands is expected, an evanescent connection between the bands should appear to preserve the overall number of bands at any frequency. In order to observe this, we fix our attention in the bandgap for the complete periodic system. Figure \ref{fig:FFTfields_EPWE}a shows the complex band structure for this range of frequencies. Let us analyze the first symmetric (respectively, antisymmetric) guided band labelled as \textit{sr1} (\textit{ar1}) in Figure \ref{fig:FFTfields_EPWE}a. Note that this band is connected to the second symmetric (antisymmetric) band, \textit{sr2} (respectively, \textit{ar2}), by means of a complex band with non zero imaginary part \textit{si1} (\textit{ai1}). It is worth noting that both the maximum value of the band \textit{sr1} (\textit{ar1}) and the minimum value of the \textit{sr2} (\textit{ar2}) are at point X. Thus the complex band that connects these two symmetric (antisymmetric) real bands presents a constant value of the real part whereas the imaginary part follows a connecting path between the two real bands preserving the symmetry of the band. 

%Figure \ref{fig:FFTfields_EPWE}b2 (\ref{fig:FFTfields_EPWE}c2) shows the symmetric (antisymmetric) acoustic field for a mode in this complex band $si1$ ($ai1$) marked with a red square. The acoustic field has been obtained from the Fourier transform of the eigenvectors for the corresponding frequency. Due to the complex value of $k$, the acoustic field shows an evanescent behaviour decaying along the waveguide. 
In the range of frequencies between the two symmetric (antisymmetric) bands, PWE predicts that due to the antisymmetry (symmetry) of the acoustic field, stop (guiding) bands are expected. Here, between the symmetric bands $sr1$ and $sr2$, an evanescent symmetric bands, $si1$, is excited whereas between the two antisymmetric bands $ar1$ and $ar2$, a symmetric real band is excited, $sr2$. Thus, the interpretation of EPWE gives a complete picture of the physics of the problem. The stop bands can be interpreted as ranges of frequencies where symmetric evanescent modes are excited; the decay rate of the mode inside the SCW being linked to the imaginary part of the complex wave vector. 

\begin{figure*}[hbt]
\begin{center}
\includegraphics[width=162mm,height=75mm,angle=0]{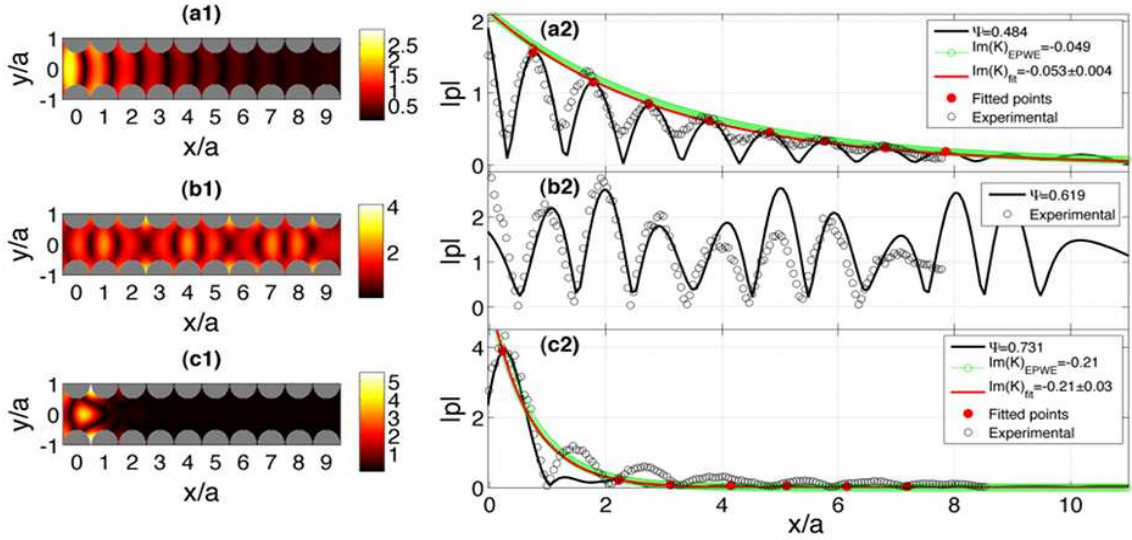}
\end{center} %\begin{quote}
\caption{(Color online) Multiple scattering analysis of the excited modes in a waveguide made by removing one row of scatterers located at $y/a=0$ in a SC constituted of 10$\times$10 cylinders. (a1)-(a2), (b1)-(b2), (c1)-(c2) show the acoustic field inside the waveguide and the longitudinal cut at $y=0$ (black continuous line) for the frequencies 0.484, 0.619 and 0.713 respectively. Red line shows the exponential fit of the red circles that represents the decay of the mode. Green line represents the decay predicted using the $Im(K)$ from EPWE. Open black circles represent the measurements in the experimental set up (see Section \ref{sec:setup}).}
 \label{fig:MST}
%\end{quote} 
\end{figure*}

To interpret these results we have selected two different modes at two different frequencies, one at $\Psi_1=0.484$ ($K_1=1/2+0.049\imath$) and the other one at $\Psi_2=0.619$ ($K_2=0.35$). The analyzed modes are marked in the band structure in Figure \ref{fig:FFTfields_EPWE}a with red squares. For the analysis we only select the first complex band because the complex bands with high values of $Im(K)$ do not contribute to the transmission properties of the SCW. Figures \ref{fig:FFTfields_EPWE}b1 and \ref{fig:FFTfields_EPWE}b2 (\ref{fig:FFTfields_EPWE}c1 and \ref{fig:FFTfields_EPWE}c2) show the acoustic field of the mode for the real and complex eigenvalue at the frequency $\Psi_1$ ($\Psi_2$) respectively. The acoustic field has been obtained from the Fourier transform of the eigenvectors for the corresponding frequency. We would like to note that the real mode at the band $ar1$ ($sr2$) is exactly the same as the predicted using PWE with a value of $K=0.33$ ($K_2=0.35$) presenting antisymmetric (symmetric) acoustic field. Oppositely, the complex mode only predicted by EPWE at the band $si1$ ($ai1$) is characterized by a value of the Bloch vector $K_1=1/2+\imath 0.049$ ($K=1/2+\imath 0.15$) with a symmetric (antisymmetric) evanescent acoustic field. Thus, because of the symmetry of the SCW, only the modes with symmetric acoustic field are excited. Therefore for the frequency $\Psi_1$ the stop band is characterized by the evanescent mode shown in Figure \ref{fig:FFTfields_EPWE}b2 whereas at frequency $\Psi_2$ the symmetric propagating mode shown in Figure \ref{fig:FFTfields_EPWE}c1 is expected.  

The connection between the bands $sr2$ and $sr3$ is different than in the previous cases. The maximum value of the band $sr2$ is at $\Gamma$ point whereas the minimum value of the band $sr3$ is at X point. Then, the evanescent connection between these bands presents both a real part that crosses the path $\Gamma$X, $sre$, and an imaginary part, $sie$, connecting the bands in the imaginary part. This connection produces an additional band that has never been predicted by PWE. To analyze this situation we have selected two modes at frequency $\Psi_3=0.731$, %For this frequency, two modes could contribute to the transmission properties, 
one with real value of the Bloch vector, $K=0.44$, at band $ar2$ and another one with complex value of the Bloch vector $K_3=0.33+\imath0.21$, being the real part of $K$ in the band $sre$ and the imaginary part in the band $sie$. %These modes are marked with coloured squares in Figure \ref{fig:FFTfields_EPWE}a. 
The acoustic fields of the real and the complex modes are shown in Figure \ref{fig:FFTfields_EPWE}d1 and \ref{fig:FFTfields_EPWE}d2 respectively. One can observe that the acoustic field corresponding to the real mode, shown in Figure \ref{fig:FFTfields_EPWE}d1, is antisymmetric and cannot be excited in the SCW. However the evanescent field produced by the complex mode, Figure \ref{fig:FFTfields_EPWE}d2, shows a symmetric acoustic field and, as a consequence, it is an excitable mode in the SCW.

The EPWE predictions allow us to affirm that the two stop bands in the SCW (grey areas in Figure \ref{fig:IL_MST}b) are characterized by two evanescent symmetric modes. The first stop band is characterized by an evanescent band that presents lower values of the imaginary part than in the case of the second stop band. This feature explains both the dependence of the IL spectra on the number of rows and the stability of the attenuation level of the second peak shown in Figure \ref{fig:IL_MST}c. However, due to the assumption of periodicity, the EPWE predictions are for an infinite SCW. In order to both check and clarify the previous results we have calculated the acoustic fields in a real waveguide made from a complete SC of $11\times 11$ rows in which we have created a linear defect in the central column (columns are defined in the direction of incidence).

Figure \ref{fig:MST} shows the MST calculation for the three frequencies previously analyzed: (a) $\Psi_1=0.484$, (b) $\Psi_2=0.619$ and (c) $\Psi_3=0.731$. Figures marked with 1's shows the absolute value of the acoustic pressure field inside the waveguide, whereas the Figures labelled with 2's show the longitudinal cut of the field inside the waveguide for $y/a=0$. The MST calculation predicts acoustic fields inside the waveguide similar to the symmetric ones obtained from the Fourier transform of the eigenvectors of the EPWE for the analyzed frequencies. This means that the excited acoustic field in the SCW corresponds to the first symmetric mode predicted by EPWE. Thus, the mode at $\Psi=$0.619 should propagate through the waveguide whereas the modes at $\Psi=$0.484 and $\Psi=$0.731 should be evanescent and the decay should be comparable with the value of the imaginary part of the $K$. 

To analyze both the propagating and the evanescent behaviour of modes we study the acoustic field in the longitudinal cut at $y/a=0$. Figure \ref{fig:MST}b2 represents the cut corresponding to the propagating mode at $\Psi=0.619$. One can observe that the field does not present any decay and the wave is guided through the waveguide. Oppositely, Figures \ref{fig:MST}a2 and \ref{fig:MST}c2 represent the cuts for the evanescent modes at 0.484 and 0.731. These cuts show clearly the decay produced because of the evanescent nature of the excited mode. The experiments (open black circles) are in good agreement with the evanescent behaviour of the modes. 

In order to check that effectively the excited modes in the stop bands corresponds to the evanescent mode obtained using EPWE, we have fitted the decay of the modes to an exponential $ae^{-bx}$. The fit of the mode at $\Psi=0.484$  using the maximum points shown in Figure \ref{fig:MST}a2 with red circles, gives a value of the decay rate $Im(K)_{fit}=0.053\pm0.004$. This value is in complete agreement with that predicted by EPWE, $Im(K)=0.049$. On the other hand, analogous procedure gives a decay rate for the mode at $\Psi=0.731$ equal to $Im(K)_{fit}=0.21\pm0.03$ in complete agreement with $Im(K)=0.21$ (Figure \ref{fig:MST}c2) predicted by EPWE. %Thus, the value of  $Im(K)$ takes an important role in the attenuation properties of the modes, in fact the attenuation level for a related number of rows is determined by this parameter.

\section{Evanescent coupling and level repulsion states in SCW}
\label{sec:coupling}

The interaction between stubs or point defects with waveguides has been extensively analyzed in the literature \cite{Vasseur05, Khelif02, Benchabane05}. The presence of the stub or a defect in a SCW alters drastically the transmission spectrum. Although the general characteristics of the transmission properties of the SCW are preserved once the stub or the defect is created in the waveguide, narrow attenuation peaks appear in the guided bands occurring at the resonance modes of the stub. The stub modes are evanescent and strongly localized by the surrounding periodic structure.

\begin{figure*}[hbt]
\begin{center}
\includegraphics[width=165mm,height=80mm,angle=0]{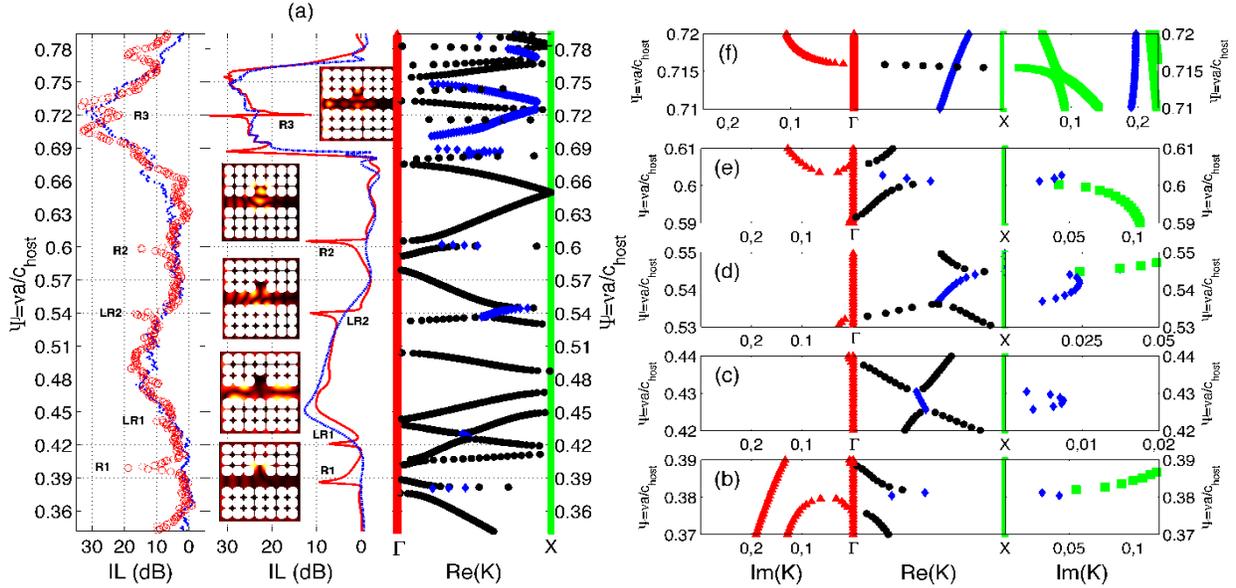}
\end{center} %\begin{quote}
\caption{(Color online) Theoretical and experimental analysis of the stubbed waveguide using EPWE and MST. (a) IL spectrum evaluated at the end of the waveguide. In the central panel blue continuous line (red dashed line) represent the MST predictions and in the left panel  blue dots (open red circles) represent the experimental results of the SCW (stubbed SCW). Peaks labelled as R (LR) represent the effect of the resonances of the stub (effect of the level repulsion between bands with different symmetry). The right panel represents the real part of the complex band structure for the SCW. (b)-(f) show the details of both the hybridization of the resonances and the guide or evanescent modes and the repulsion between bands (see text).}
 \label{fig:stubbed_waveguide}
%\end{quote} 
\end{figure*}  

In this Section, we investigate both theoretically and experimentally the occurrence of the effects produced by a stub in a SCW. Stubs are generated in this work by removing a scatterer in the side-wall of the waveguide. We observe two different effects due to the stub. On one hand, EPWE predicts attenuation dips due to the interaction of both the guided and the evanescent modes with the resonances of the stub. This result constitutes a proof of the importance of the $k(\omega)$ methods because they predict the evanescent bands and, as a consequence, the interaction between the evanescent modes and the resonances of the stub. The attenuation peaks in Figure \ref{fig:stubbed_waveguide}a marked with R (resonance) correspond to this first effect and they will be discussed in detail later.

On the other hand, the presence of the stub changes the symmetry of the system, therefore antisymmetric modes can be excited even by symmetric longitudinal incident waves. This produces the appearance of repulsion states between symmetric and antisymmetric bands. In this work we observe that due to the local breaking of the symmetry because of the presence of the stub, a repelling state appears between symmetric and antisymmetric bands accompanied by a transfer of the symmetry of the modes from one band to the other. Chen \textit{et al.} showed that depending on the symmetry of the system with respect to the incident direction of the incident wave, shear-horizontal modes either couple or not with the Lamb wave modes which are polarized in the sagittal plane. The coupling can be observed by a splitting between bands in the band structure and by a transfer of the symmetry from one band to the other. In the case in which no coupling occurs, the symmetric Lamb waves band simply crosses of the shear-horizontal band. Similar results have been recently observed in 1D phononic crystals \cite{Bavencoffe09b, Bavencoffe09a}, and anisotropic phononic crystals \cite{Achaoui10}. We observe in this work a new interpretation: The repelled bands are connected by an evanescent mode which is the responsible of the transfer of the symmetry from one band to another. The attenuation peaks in Figure \ref{fig:stubbed_waveguide}a marked with LR (level repulsion) correspond to this second effect and they will be discussed in detail later.

The real part of complex band structures of the stubbed waveguide are also shown in the right panel of the Figure \ref{fig:stubbed_waveguide}a. One can observe the correspondence between the attenuation peaks, R and LR, predicted by MST (central panel of Figure \ref{fig:stubbed_waveguide}a) and the singular points of the band structure. Left panel of Figure \ref{fig:stubbed_waveguide}a shows the experimental data (see Section \ref{sec:setup} for the details of the experimental set up) in good agreement with the theoretical predictions. Let us study the nature of each of these attenuation peaks.

%\subsection{Stub resonances}
\subsection{Coupling between guided and evanescent modes with stub resonances}
\label{sec:waveguide_coupling}
\begin{figure}[hbt]
\begin{center}
\includegraphics[width=85mm,height=35mm,angle=0]{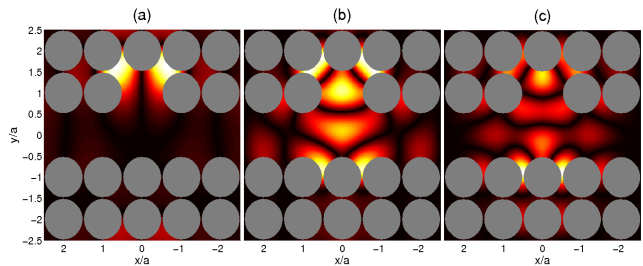}
\end{center} %\begin{quote}
\caption{(Color online) Absolute value of the acoustic field for the stub resonances. (a) R1, $\Psi_1$=0.3812. (b) R2, $\Psi_2$=0.605. (c) R3, $\Psi_3$=0.72.}
 \label{fig:stub_resonances}
%\end{quote} 
\end{figure}

The stub acts as a resonant cavity in the waveguide producing flat bands due to the localized mode in the stub. If this resonance occurs at the same frequency as a guiding or evanescent bands, then the hybridization of the two bands produces an attenuation peak in the spectrum. Resonances of the stub, calculated using EPWE at point X, can be observed in Figure \ref{fig:stub_resonances}. It is worth noting that the frequencies of these resonances coincide with two symmetric guiding bands and one evanescent band of the waveguide. 

Points R1 ($\Psi=0.3812$) and R2 ($\Psi=0.605$) in Figure \ref{fig:stubbed_waveguide}a coincide with the hybridization of the guiding bands and the resonance bands of the stub (flat band). A detailed image of the hybridization effect at these frequencies can be seen in Figures \ref{fig:stubbed_waveguide}b and \ref{fig:stubbed_waveguide}e. These hybridizations explain the attenuation peaks in the IL spectrum labelled as R1 and R2 in Figure \ref{fig:stubbed_waveguide}a in good agreement with both the analytical predictions and the experimental results. The shape of the attenuation peaks assembles the Fano-like resonances previously reported \cite{Goffaux03}. The slight disagreements between the experimental data and the theoretical  predictions can be due to the sensitivity of the experimental data to the size of the stub.

\begin{figure*}[hbt]
\begin{center}
\includegraphics[width=165mm,height=80mm,angle=0]{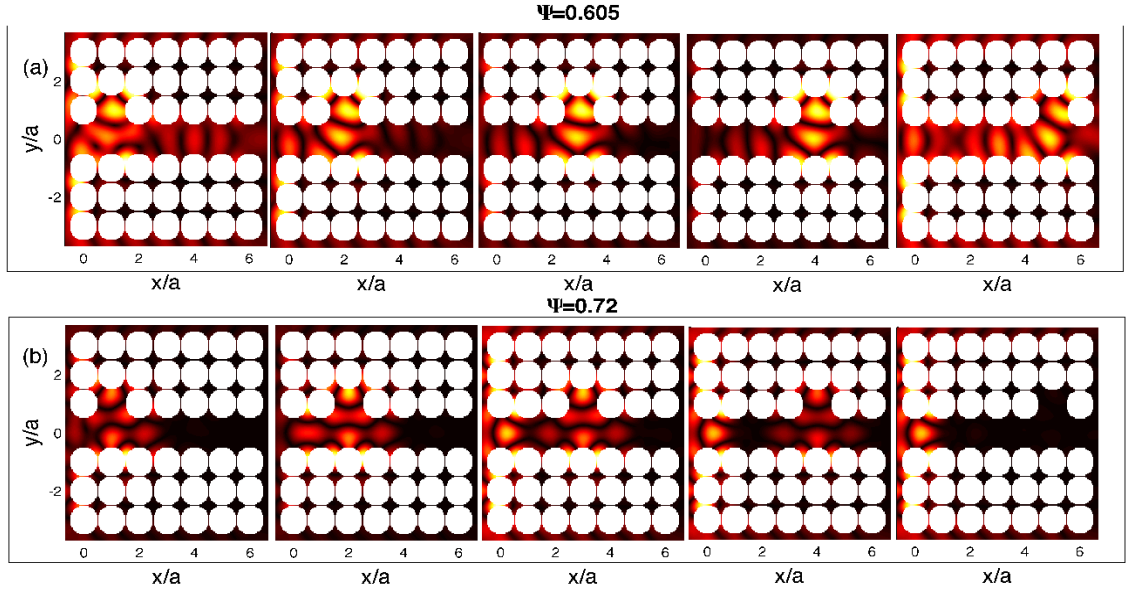}
\end{center} %\begin{quote}
\caption{(Color online) Coupling of guided and evanescent modes with stub resonances. (a) Coupling of the resonance at $\Psi=0.605$ with the guided mode at the same frequency. (b) Coupling of the resonance at $\Psi=0.72$ with the evanescent mode at the same frequency.}
 \label{fig:coupling}
%\end{quote} 
\end{figure*}

The interaction between the guided mode in the SCW and the resonance of the stub at point R2 is analyzed in Figure \ref{fig:coupling}a. The calculations have been carried out using the MST. We study the propagation of waves at frequency $\Psi=0.605$ inside a SCW $7a$ long with a stub. The position of the stub has been changed all along the SCW in order to observe that the guided mode excites the resonance of the stub independently of the distance between the site of the stub and the entrance of the SCW (left side of the SCW). The wave penetrates in the waveguide exciting the symmetric guided mode. Once the guided mode finds the stub, the resonance of the stub is excited, localizing the field in that region. One can compare the field around the stub in Figure \ref{fig:coupling}a, calculated using MST, with the acoustic field of the resonance of the stub in Figure \ref{fig:stub_resonances}b calculated using EPWE. It is worth noting that the resonance of the stub is excited practically with the same intensity in all the locations of the stub due to the low losses of the guided mode. Analogous effect occurs for the point R1 ($\Psi=0.3812$). 

Point R3 ($\Psi=$0.72) corresponds to the interaction of an evanescent mode with the resonance of the stub at this frequency. The detailed band structure for this frequency is shown in Figure \ref{fig:stubbed_waveguide}f. Note that for this case no hybridization is produced and two bands can be excited. Once the wave penetrates in the SCW the evanescent mode is excited and, if the stub is as close as the evanescent mode can travel through the SCW, the stub mode should be excited. Note that these results are not predicted by the classical methods, therefore they cannot be explained with the PWE, but EPWE gives the correct understanding of the phenomenon.

MST simulations have been carried out in order to observe this coupling between the evanescent mode and the stub resonances at $\Psi=$0.72. Figure \ref{fig:coupling}b shows five stubbed waveguides with increasing distances between the sites of the stub and the entrance of the SCW. In all cases, the excited evanescent mode presents a symmetric acoustic field equivalent to that predicted by EPWE in Figure \ref{fig:FFTfields_EPWE}d2. Once this mode arrives to the stub, it excites the stub resonance. However it is worth noting that the mode is evanescent and, as a consequence, the intensity of the localized modes in the stub should decrease as the stub is far from the entrance of the SCW. The MST calculations predicts this behaviour. For the last case, the distance between the stub and the entrance of the SCW is larger than the penetration distance of the evanescent mode and, as a consequence, any stub resonance can be excited.

\subsection{Level repulsion states}
\label{sec:level_repulsion}
In addition to the attenuation peaks due to the resonances, the IL spectrum of the stubbed waveguide (Figure \ref{fig:stubbed_waveguide}a) predicts two additional attenuation peaks referred to as LR1 ($\Psi=$0.429) and LR2 ($\Psi=$0.544). These peaks have different phenomenology than R1, R2 or R3 peaks which are due to the interaction between the guided or evanescent modes with the stub resonances (see previous Section). Oppositely, the LR1 and LR2 peaks correspond to crossing points in the band structure shown in Figure \ref{fig:stubbed_waveguide}b. 

\begin{figure*}[hbt]
\begin{center}
\includegraphics[width=165mm,height=80mm,angle=0]{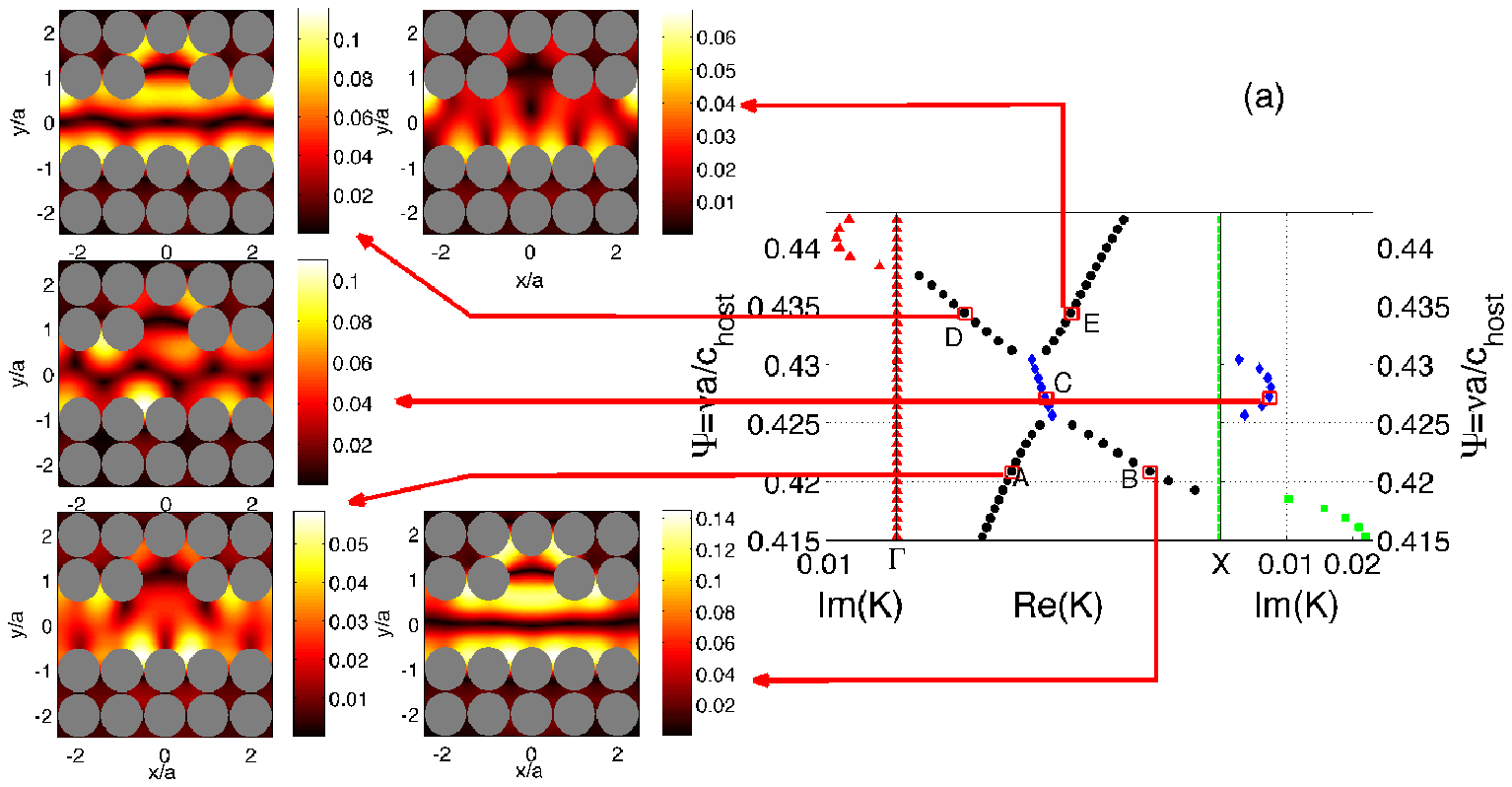}\\
\includegraphics[width=165mm,height=80mm,angle=0]{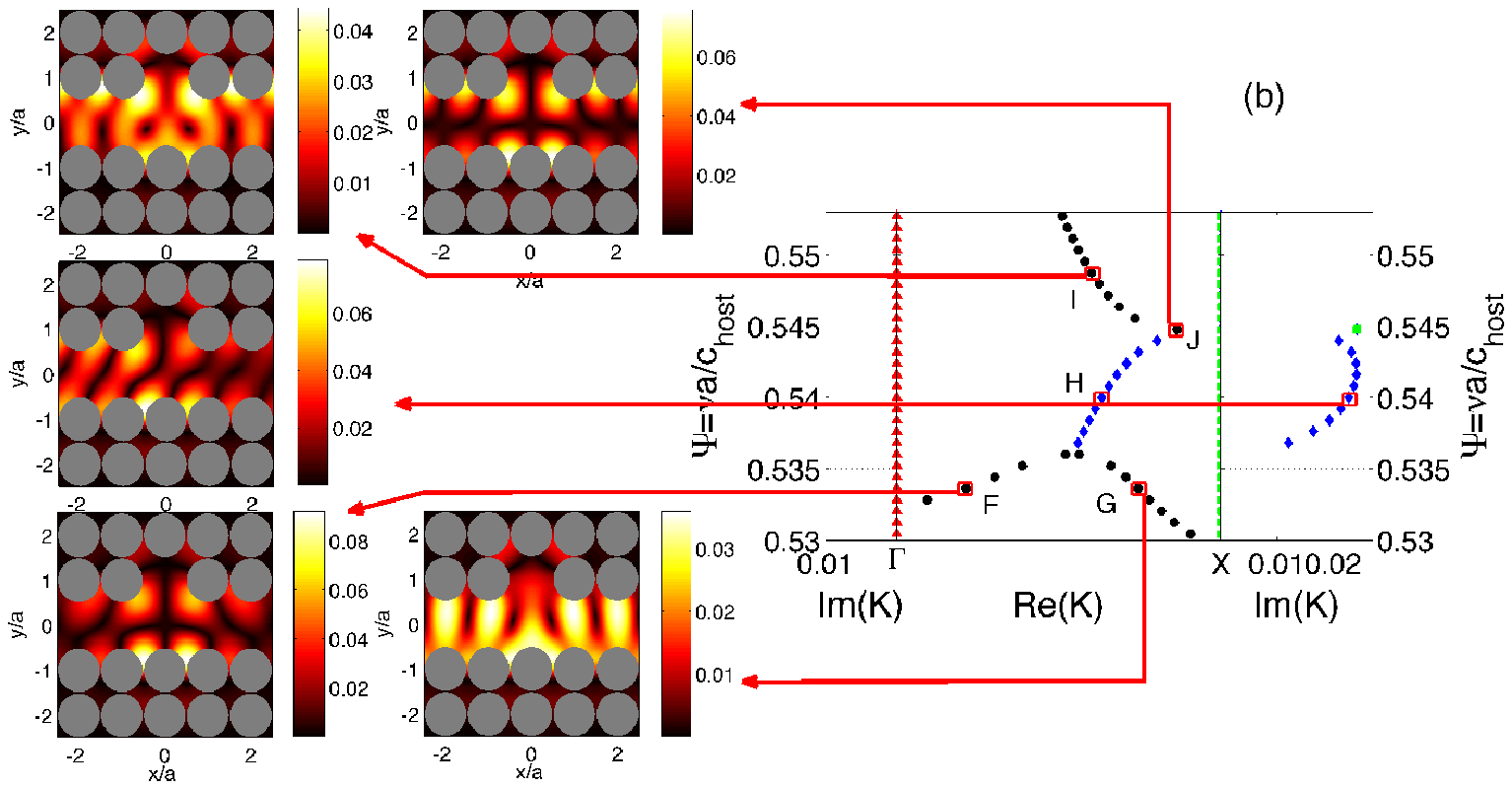}
\end{center} %\begin{quote}
\caption{(Color online) Analysis of the symmetry transfer between the repelled bands in stubbed waveguides. The analysis in (a)  and (b) correspond to the repulsion states shown in Figure \ref{fig:stubbed_waveguide}f and \ref{fig:stubbed_waveguide}e respectively. The acoustic fields are represented using the absolute value of the pressure obtained from the Fourier transform of the eigenvectors for the eigenvalues marked with red squares.}
 \label{fig:symmetry_transfer}
%\end{quote} 
\end{figure*}

Very recent works \cite{Achaoui10} have shown several kinds of crossing points in anisotropic phononic crystals. As a general rule for these systems, the bands involved in the crossing point, or level repulsion state, exchange their polarization, so that the polarization remains a continuous function of the wave vector $\vec{k}$. For the case of SC in which only longitudinal polarization is allowed, one can observe symmetric and antisymmetric acoustic fields. If the SC is symmetric with respect to the incident acoustic field, only symmetric acoustic fields will be supported. However, if the system is non symmetric, then both symmetric and antisymmetric fields can be supported. The stubbed SCW is an example of a non symmetric system.

In our case, the points LR1 and LR2 correspond to level repulsion states in the band structures that introduce a local band gap. The level repulsion state for these cases appears due to the crossing between an antisymmetric and a symmetric bands in a similar way as in phononic crystals thin plates \cite{Chen08, Bavencoffe09b, Bavencoffe09a}. Figures \ref{fig:stubbed_waveguide}c and \ref{fig:stubbed_waveguide}d represent detailed views of the band structures calculated using EPWE for the points LR1 and LR2 respectively. One can see that, in both cases, the repelled bands are connected by means of an additional band, which is evanescent. Thus, EPWE does not predict a repulsion level state but an evanescent connection between bands.

In order to analyze in more details the phenomenology involved in these level repulsion states, we have analyzed several modes for different frequencies all along the bands involved in the repulsion. Figure \ref{fig:symmetry_transfer} shows the acoustic fields obtained from the Fourier transform of the eigenvectors for several modes in the repelled bands. The analyzed modes are marked with red squares which are connected using red arrows with their corresponding acoustic fields.

One can observe that the modes in the repelled bands present a completely real eigenvalue, $k$, whereas the connection bands have complex eigenvalue. The repelled bands are predicted by EPWE and they are exactly the same as the ones predicted by the PWE, however the difference appears in the evanescent connection between the repelled bands.

First, we fix our attention on the acoustic fields in the repelled bands. One can observe in Figure \ref{fig:symmetry_transfer} that for each band, depending on the position with respect to the evanescent connection, different symmetries of the field can be excited. For example points A and B that belong to the same band have different symmetries. A is on the left of the evanescent connection and B is on the right of the evanescent connection. The acoustic field corresponding to the point A is mainly symmetric. However the acoustic field for the point B is completely antisymmetric. Then, inside the band, there is an exchange of the symmetry, so that the symmetry remains a continuous function of the wave vector $\vec{k}$ for each band. The same situations appear for the couple of modes D-E, F-G, an I-J.

The repelled bands are connected by an evanescent band which is the responsible of the exchange of symmetry between bands. The acoustic fields of the modes inside the evanescent connection have a mixed symmetry. We have studied the modes C and H in the evanescent connections. First of all one can see the mixed symmetry of the acoustic field for the modes C and H. Moreover, one can observe the evanescent behaviour of the acoustic field inside the guide with a low decay rate indicating the small value of the imaginary part of the eigenvalue, $k$. It is worth to noting that the acoustic field predicted by MST for the frequencies corresponding to the modes C and H is fairly close to the one predicted by EPWE (Figure \ref{fig:coupling}). The MST predictions can be seen in the Figure \ref{fig:stubbed_waveguide}a.

\subsubsection{Symmetric and antisymmetric stubbed SCW.}

\begin{figure*}[hbt]
%\begin{center}
\includegraphics[width=80mm,height=50mm,angle=0]{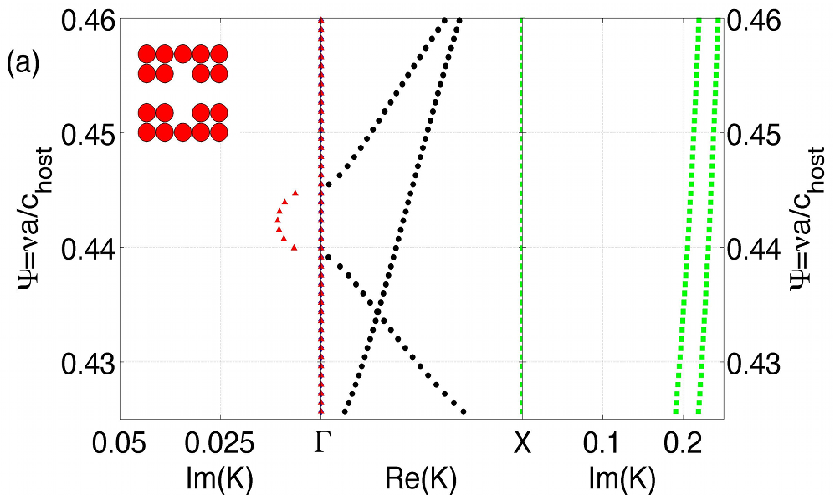}
\includegraphics[width=80mm,height=50mm,angle=0]{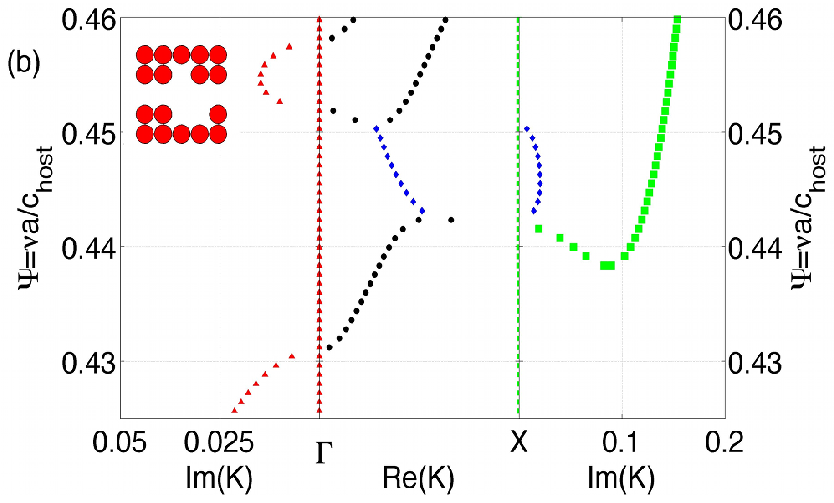}
%\end{center} %\begin{quote}
\caption{(Color online) Complex band structures of two stubbed SCW, one symmetric (a) and the other one antisymmetric (b). The supercells for the EPWE calculations are shown in the insets. The evanescent connection between the repelled bands is marked with a blue diamonds for the case of the antisymmetric stubbed SCW.}
 \label{fig:symmetric_antisymmetric_stub}
%\end{quote} 
\end{figure*}

Following the previous results, the repulsion states in SC should appear in systems presenting some degree of antisymmetry, as for example a stubbed SCW. In the benefit of the clarity of the previous results, we have analyzed two additional systems, a symmetric and another one with a degree of antisymmetry. Figure \ref{fig:symmetric_antisymmetric_stub}a shows the complex band structure for a completely symmetric system with a symmetric stub as shown in the inset. Figure \ref{fig:symmetric_antisymmetric_stub}b shows the complex band structure for the SCW with an antisymmetric stub which supercell is also shown in the inset.

One can observe that the band structure of the symmetric stubbed SCW presents a crossing point but the bands are not connected, i.e., they are independent and only the symmetric one will contribute to the transmission properties of the system. However, for the case of the antisymmetric stubbed SCW one can observe again the evanescent connection between the repelled bands (blue diamonds). 

\section{Concluding remarks}
\label{sec:conclusions}
The complex band structures obtained using the EPWE ($k(\omega)$ method) show additional bands never revealed by the classical $\omega(\vec{k})$ methods. In this work we have shown both theoretically, with two independent theoretical techniques (EPWE and MST), and experimentally, making use of 3DReAMS, the interpretation of these evanescent additional bands. Due to the conservation of the overall number of bands for a determined frequency, interesting interpretations of the deaf bands and the level repulsion states have been obtained depending on the symmetry or not of the system. In the ranges of frequencies where a deaf band is traditionally predicted an evanescent mode with the excitable symmetry appears changing drastically the transmission properties of the system. On the other hand, we have interpreted, without loss of generality, the level repulsion between symmetric and antisymmetric bands in antisymmetric sonic crystals stubbed waveguides as the presence of an evanescent mode connecting both bands. These evanescent modes explain both the attenuation produced in this range of frequencies and the transfer of symmetry from one band to the other. Thus, the additional bands of EPWE are not an artifact of the calculation but a real physical behaviour can be attributed to them and as a consequence, the evanescent modes should be considered in the design of the systems based on periodicity. 

\ack
VRG and LMGR would like to thank the hospitality and the facilities provided by the Institut d'Electronique, de Micro\'electronique et de Nanotechnologie (IEMN UMR CNRS 8520). LMGR would like to thank the  Universidad Polit\'ecnica de Valencia for the grant ``Programa de Apoyo a la Investigaci\'on y Desarrollo (PAID-00-11)''. VRG is grateful for the support of ``Programa de Contratos Post-Doctorales con Movilidad UPV (CEI-01-11)''. This work was supported by MCI Secretar\'ia de Estado
de Investigaci\'on (Spanish government) and the FEDER funds, under grant MAT2009-09438.

\end{document}